%% file: main_file.tex
\documentclass[10pt,aps,pra,twocolumn,groupedaddress]{revtex4-1}

\usepackage{amsmath}
\usepackage{graphicx}
\usepackage{color}
\usepackage{epstopdf}
\usepackage{hyperref}

\newcommand{\e}{\mathrm{e}}
\newcommand{\rmi}{\mathrm{i}}
\newcommand{\W}{\mathcal{W}}
\newcommand{\JT}{\mathcal{A}}
\newcommand{\JS}{\mathcal{A}}
\newcommand{\s}{\mathrm{s}}
\renewcommand{\i}{\mathrm{r}}
\newcommand{\p}{\mathrm{p}}
\newcommand{\q}{\mathrm{q}}
\renewcommand{\c}{\mathrm{c}}

\newcommand{\A}{\hat{a}}
\newcommand{\Ad}{\hat{a}^{\dagger}}
\newcommand{\vac}{\mathrm{vac}}

\newcommand{\ket}[1]{\left | #1 \right \rangle}

\renewcommand{\H}{\hat{H}}
\newcommand{\HC}{\,\mathrm{h.c.}}
\newcommand{\ud}{\,\mathrm{d}}
\newcommand{\m}{\hat{m}}

\newcommand{\M}{\hat{M}}

\newcommand{\RR}{\mathcal{R}}
\newcommand{\R}{\mathrm{R}}
\newcommand{\mean}[1]{ \left \langle #1 \right \rangle}
\renewcommand{\vec}[1]{\mathbf{#1}}
\newcommand{\qqquad}{\quad\qquad}
\newcommand{\cc}{\mathrm{c.c.}}
\newcommand{\ip}[2]{\langle #1 | #2 \rangle}
\renewcommand{\a}{\hat{A}}
\newcommand{\ad}{\hat{A}^\dagger}

\newcommand{\normord}[1]{:\mathrel{#1}:}

\definecolor{col1}{RGB}{0,0,115}
\definecolor{col2}{RGB}{138,0,0}
\definecolor{col3}{RGB}{0,0,0}

\begin{document}


\title{Photon-pair generation by non-instantaneous spontaneous four-wave mixing}


\author{Jacob G. Koefoed}
\email[]{jgko@fotonik.dtu.dk}
\author{Jesper B. Christensen}
\author{Karsten Rottwitt}
\affiliation{Department of Photonics Engineering, Technical University of Denmark, 2800 Kongens Lyngby, Denmark}


\date{\today}

\begin{abstract}
We present a general model, based on a Hamiltonian approach, for the joint quantum state of photon pairs generated through pulsed spontaneous four-wave mixing, including nonlinear phase-modulation and a finite material response time. For the case of a silica fiber, it is found that the pair-production rate depends weakly on the waveguide temperature, due to higher-order Raman scattering events, and more strongly on pump-pair frequency detuning.
From the analytical model, a numerical scheme is derived, based on the well-known split-step method. This scheme allows computation of joint states where nontrivial effects are included, such as group-velocity dispersion and Raman scattering. In this work, the numerical model is used to study the impact of the non-instantaneous response on the pre-filtering purity of heralded single photons. We find that for pump pulses shorter than 1 ps, a significant detuning-dependent change in quantum-mechanical purity may be observed in silica. This shows that Raman scattering not only introduces noise, but can also drastically change the spectral correlations in photon pairs when pumped with short pulses.
\end{abstract}

\pacs{}

\maketitle

\section{Introduction}
Preparation and distribution of single-photon states is vital to many emerging quantum technologies such as quantum communication~\cite{Gisin2007,Tanzilli2005,Kimble2008}, quantum cryptography~\cite{Ekert1991,Gisin2002} and linear optical quantum computation~\cite{Knill2001,Ladd2009}. A promising way to prepare single photons is by nonlinear optical processes that generate photon pairs such as spontaneous parametric down conversion~\cite{Tanzilli2001,Banaszek2001,URen2004} or spontaneous four-wave mixing (SpFWM) in e.g. dispersion-shifted fibers~\cite{Li2005}, photonic crystal fibers~\cite{Sharping2004, Rarity2005,Cohen2009} or silicon waveguides~\cite{Sharping2006, Xiong2011}. Even though such nonlinear pair-production processes are inherently probabilistic, this can in many cases be compensated for by \textit{heralding}~\cite{Fasel2004,McMillan2009}, whereby detection of one photon in the pair implies the existence of the other. If reliable photon production is required, near-determistic behaviour can in principle be achieved by multiplexing of such heralded probabilistic sources~\cite{Ma2011, Collins2013, Xiong2016}, \par
SpFWM processes posses a large number of tunable parameters, allowing great flexibility in the choice of single-photon wavelengths as well as the temporal and spectral properties of generated photon pairs~\cite{Garay-Palmett2007}. Importantly, this allows single photons to be generated at communication wavelengths and in fiber, compatible with conventional communication systems~\cite{Fiorentino2002}. This results in small losses, which is ideal for quantum communication purposes. The flexible nature of SpFWM can also be used to generate heralded single photons with a high quantum-mechanical purity~\cite{Grice1997}, without the use of extensive filtering~\cite{Halder2009,Soller2011,Clark2011}. This property is crucial for applications in linear optical quantum computing based on photon interference~\cite{Hong1987}, which relies on photon indistiguishability~\cite{Walmsley2005, Mosley2008}. The use of non-degenerate pulsed pumps offers even greater flexibility and has been proposed for generating photons of very high pre-filtering purity~\cite{Christensen2016}.\par
It is well known that spontaneous Raman scattering (SpRS) represents a significant noise source in many experimental realizations of photon-pair generation through SpFWM~\cite{Clark2012}. For this reason there have been several studies, using the Heisenberg picture, considering the impact of Raman noise on photon statistics for a continuous or narrowband pump and narrow spectral filtering in the context of SpFWM~\cite{Lin2007, Silva2012, Silva2013, Voss2004, Voss2006}. However, pulsed pumping is of significant practical interest, due to better noise performance~\cite{Dyer2009} and the ability to multiplex several sources~\cite{Xiong2016}. Additionally, most schemes for producing unfiltered pure heralded photons rely critically on the use of broadband pumps~\cite{Grice1997,Garay-Palmett2007}. In the case of photonic crystal fibers, which has been one of the primary platforms for demonstration of pure single-photon generation~\cite{Cohen2009, Francis-Jones2016}, it is desirable to use short pump pulses and shorter fiber lengths to reduce the impact of fabrication imperfections along the fiber length~\cite{Cui2012}. However, little attention has been devoted to study the impact on temporal and spectral correlations of generated photon-pairs under a non-instantaneous nonlinear material response, which may become important for short pump pulses.\par
In this work, we adopt an interaction picture formalism to study the impact of photon-phonon interactions on the spectral and temporal joint state of photon pairs generated through SpFWM. We present a general analytic pulsed-source model, including a non-instantaneous nonlinear response as well as nonlinear phase modulation (NPM). We use the model to characterize source performance by investigating the pre-filtering photon statistics and two-photon joint amplitude. We also present a numerical split-step scheme for efficiently computing the joint state. The numerical model provides a strong and versatile tool for simulation of realistic systems, incorporating all effects of interest that cannot simultaneously be included in analytical models, such as nonlinear phase modulation, higher-order dispersion and Raman scattering. The numerical model is easily generalized to include additional effects that may be desired, as well as to non-degenerate pulsed setups.

\section{Theory}

\begin{figure}[ht]
\centering
\def\svgwidth{1\linewidth}
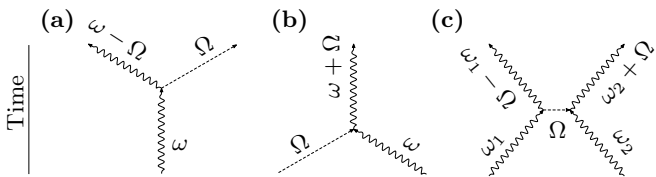
\caption{Feynman diagrams where wavy lines represent photons and dashed lines represent phonon excitations. \textbf{(a)} Stokes scattering where a photon is scattered into a lower energy photon while creating a phonon.  \textbf{(b)} Anti-stokes scattering where a photon is scattered into a higher energy photon while absorbing a phonon. \textbf{(c)} Higher-order diagram from combination of \textbf{(a)} and \textbf{(b)} where a phonon mediates the scattering of two incoming photons into two outgoing photons with one gaining and one losing energy corresponding to the phonon mode.}
\label{Feynman}
\end{figure}

We consider a SpFWM process in which a pump field, denoted by subscript $\p$, propagates through a $\chi^{(3)}$-nonlinear medium. In the nonlinear SpFWM process, two pump photons may then be spontaneously annihilated to create a signal ($\s$) and an idler ($\i$) photon. The central frequencies, $\omega_{\s 0}$ and $\omega_{\i 0}$, of the created photon pair are determined by energy and momentum conservation such that
\begin{subequations}
\begin{align}
2\omega_{\p 0} -\omega_{\s 0} - \omega_{\i 0} &= 0,\\
2\beta_{\p 0} -\beta_{\s 0} - \beta_{\i 0} &= 0,
\end{align}
\end{subequations}
where $\beta_{j0}$ for $j = \p,\s,\i$ is the propagation constant at the central frequency $\omega_{j0}$. Due to nonlinear phase-modulation, which is included in this analysis, the central frequencies of the generated field are slightly different from $\omega_{0\s}$ and $\omega_{0\i}$. The pump field is decomposed as
\begin{align}
\label{pump_field}
\vec{E}_\p = \frac{1}{2}\vec{e} F(x,y)  \sqrt{\frac{2}{n_\p\epsilon_0 c }}& \left [ A_\p(z,t)\e^{\rmi(\beta_{0\p}z - \omega_{0\p}t)} +\cc \right ].
\end{align}
Here, $F(x,y)$ describes the mode-profile, normalized such that the integral of $|F(x,y)|^2$ is unity over the waveguide cross-section and $n_\p = n(\omega_{0\p})$ is the refractive index at the pump wavelength. In this normalization, $|A_\p|^2$ represents optical power. The signal and idler fields are quantized in the following way:
\begin{align}
\vec{\hat{E}}_j(z,t) &= \frac{1}{2}\vec{e}F(x,y)\e^{\rmi(\beta_{j 0}z - \omega_{j 0}t)}\frac{1}{2\pi}  \\
&\quad\times\int \ud \omega \sqrt{ \frac{2\hbar (\omega_{j 0} + \omega)}{n(\omega_{j 0} + \omega)\epsilon_0  c }} \A_j(z,\omega) \e^{-\rmi\omega t} + \HC,\notag
\end{align}
where $j = \s, \i$ and $\A_j(z,\omega)$ ($\Ad_j(z,\omega)$) is the annihilation (creation) operator for field $j$ at the frequency $\omega$, which is relative to the central frequency of the field. We have assumed that all fields are in the same spatial mode in the waveguide although the theory could be easily extended to multiple spatial modes. If the field is spectrally narrow compared to the central frequency $\omega_{j 0}$, that is $|\omega| \ll \omega_{j0}$ for the frequency range of interest, the prefactor on the field operator inside the integral is approximately constant such that
\begin{equation}
\label{signal_field}
\vec{\hat{E}}_j(z,t) = \frac{1}{2}\vec{e}F(x,y)\e^{\rmi(\beta_{j 0}z - \omega_{j 0}t)} \sqrt{ \frac{2\hbar \omega_{j 0}}{n_j\epsilon_0  c }} \A_j(z,t) + \HC,
\end{equation}
for $j = \s,\i$, where $\A_j(z,t)$ is the field operator for field $j$, which is the inverse Fourier transform of the annihilation operator. The operator fields satisfy the equal position commutation relations
\begin{subequations}
\begin{align}
\label{commutation_field}
[\A_i(z,t),\Ad_j(z,t')] &= \delta_{ij}\delta(t-t'), & i,j &= \s,\i, \\
[\A_i(z,t),\A_j(z,t')] &= 0, & i,j &= \s,\i. 
\end{align}
\end{subequations}
We describe the system by the following interaction Hamiltonian governing spatial evolution, derived in Appendix \ref{app:el_Hamiltonian}:
\begin{align}
\label{Hamiltonian}
\H_{\mathrm{int}}(z)  &=\sqrt{\gamma_\s \gamma_\i} \int_{-\infty}^{\infty}\ud t_1\int_{-\infty}^{\infty}\ud t_2  \RR(t_1 - t_2)\notag \\
&\quad\times A_\p(z,t_1)A_\p(z,t_2)  \Ad_\s(z,t_1) \Ad_\i(z,t_2) \e^{-\rmi\Omega (t_1 - t_2)} \notag\\
&+ \int_{-\infty}^{\infty} \ud t  A_\p(z,t) \m(z,t) \\
&\quad\times\Big [\sqrt{\gamma_\s}\Ad_\s(z,t)\e^{-\rmi\Omega t}+\sqrt{\gamma_\i}\Ad_\i(z,t) \e^{\rmi\Omega t}\Big] +\HC \notag
\end{align}
Here, $L$ is the waveguide length, $\m(z,t)$ is a noise operator representing the phonon-field, $\Omega = \omega_{\i 0} - \omega_{\p 0} = -(\omega_{\s 0} - \omega_{\p 0})$ is the frequency detuning, and $\gamma_j$ is the nonlinear coefficient, given by Eq. \eqref{nonlinear_coeff_2}. The function $\RR(t)$ governs the temporal separation of creation events and is given by
\begin{equation}
\RR(t) = \frac{1}{2} \left [R(t) + R(-t)\right ].
\end{equation}
where the response function
\begin{equation}
\label{NLresponse}
R(t) =  (1-f_\R) \delta(t) + f_\R h_\R(t),
\end{equation}
has a fraction $f_\R$ from the phononic contribution to the nonlinearity and $h_\R(t)$ is the Raman response. The noise operator $\m(z,t)$ describes the coupling to the phononic noise background. The noise operator correlations in the time domain are derived in Appendix \ref{app:noise} and found to be of the form
\begin{equation}
\mean{\m(z_1,t_1) \m(z_2,t_2)} = \delta(z_1 - z_2) \mathcal{F}(t_1 - t_2),
\end{equation}
where the function $\mathcal{F}(t)$ determines the lifetime of phonon excitations. The photon-state after propagation through a waveguide of length $L$ can now be expressed through an evolution operator $\hat{U}$
\begin{equation}
\ket{\psi(L)} = \hat{U}(0,L) \ket{\vac} =\exp\left (\rmi \int_0^L \ud z \H_{\mathrm{int}}(z)\right )  \ket{\vac}.
\end{equation}
We assume in the following that time-ordering corrections~\cite{Quesada2014} can be neglected, which is a good approximation in the low gain regime~\cite{Christ2013a}, and expand the exponential in a simple Taylor series by keeping all terms where two or fewer signal/idler photons are created, or equivalently to first order in the nonlinear phase shift $\phi_\mathrm{NL} = \gamma P_\p L$. Suppressing the integral limit for convenience, with the understanding that space integrals are from $0$ to $L$ and time integrals over all time, this gives the expansion
\begin{widetext}
\begin{align}
\hat{U}&(0,L) = \mathcal{I} \notag \\
&+\rmi\sqrt{\gamma_\s \gamma_\i} \iiint \ud z_1 \ud t_1  \ud t_2 \RR(t_1-t_2)A_\p(z_1,t_1)A_\p(z_1,t_2) \Ad_\s(z_1,t_1) \Ad_\i(z_1,t_2) \e^{-\rmi\Omega (t_1 - t_2)}\notag\\
 &+\rmi\iint \ud z_1\ud t_1 A_\p(z_1,t_1) \m(z_1,t_1)\left [\sqrt{\gamma_\s}\Ad_\s(z_1,t_1)\e^{-\rmi\Omega t_1} + \sqrt{\gamma_\i}\Ad_\i(z_1,t_1)\e^{\rmi\Omega t_1}\right ]\notag \\
& -\frac{1}{2}\iiiint\ud z_1  \ud z_2 \ud t_1 \ud t_2 A_\p(z_1,t_1) A_\p(z_2,t_2) \m(z_1,t_1) \m(z_2,t_2)\label{evolution_operator} \\
&\qquad\times\Big[\gamma_\s\Ad_\s(z_1,t_1)\Ad_\s(z_2,t_2)\e^{-\rmi\Omega (t_1 + t_2)} + \gamma_\i\Ad_\i(z_1,t_1)\Ad_\i(z_2,t_2)\e^{\rmi\Omega (t_1 + t_2)}\notag \\
&\qquad \quad+ \sqrt{\gamma_\s \gamma_\i}\Ad_\s(z_1,t_1)\Ad_\i(z_2,t_2)\e^{-\rmi\Omega (t_1 - t_2)}+ \sqrt{\gamma_\s \gamma_\i}\Ad_\i(z_1,t_1)\Ad_\s(z_2,t_2)\e^{\rmi\Omega( t_1 -  t_2)}\Big] \notag \\
&+ (\text{terms with annihilation operators}) + \mathcal{O}(\phi_\mathrm{NL}^{3/2}) \notag.
\end{align}
\end{widetext}
The first term gives the vacuum state and the second term describes photon pairs produced by time-delayed SpFWM. The third term describes single photons generated by SpRS through either a Stokes or anti-Stokes scattering process. Feynman diagrams representing such events are shown in Figs. \ref{Feynman}\textbf{(a)}-\textbf{(b)}. 
The remaining terms describe higher-order phonon scattering events, which are combinations of the fundamental Stokes and anti-Stokes diagrams. These include both the trivial (and expected) unconnected combinations, which describe two independent Raman scattering events taking place at different positions in the waveguide. \newline
A more interesting higher-order diagram is the one shown in Fig. \ref{Feynman}\textbf{(c)} where two incoming photons are scattered on a single phonon-mode (at a single waveguide position). Such events can happen in two ways: In the first way, the first photon creates a (virtual) phonon with a second photon subsequently annihilating it, with probability proportional to $1+n_\mathrm{th}$ since such an interaction is possible with both the phononic ground states and excited states. In the second way, the first photon annihilates an existing phonon, with a second photon subsequently recreating it, with probability proportional to $n_\mathrm{th}$, since a thermally excited phonon has to be present prior to this interaction. These events are parametric and photon pairs created in this way are correlated similarly to the regular SpFWM pairs and thus they give a contribution to the two-photon state. Independently created Raman photons, do not contribute to the joint state since they are completely uncorrelated in their number distribution. They do in this sense not constitute a photon pair.\par
When neglecting group-velocity dispersion, the pump envelope $A_\p(z,t)$, which is treated classically and assumed undepleted, evolves only under SPM~\cite{Agrawal2013}:
\begin{equation}
\partial_z A_\p(z,t) = \rmi \gamma_\p A_\p(z,t) \int_{-\infty}^t \ud t' R(t-t') |A_\p(z,t')|^2,
\end{equation}
where we carry out all calculations in the reference frame of the pump. The solution to this is equation is
\begin{subequations}
\begin{align}
A_\p(z,t) &= A_\p(0,t) \exp\left [\rmi \theta_\p(z,t) \right ], \label{pump_evolution} \\
\theta_\p(z,t) &= \gamma_\p z \int_{0}^\infty \ud t' R(t') |A_\p(0,t-t')|^2 \label{pump_phase}.
\end{align}
\end{subequations}
We work in the interaction picture of quantum mechanics where the spontaneous scattering effects are applied to the two-photon state and the evolution of the field operators is governed only by XPM~\cite{Agrawal2013}:
\begin{align}
\label{full_field_eq}
\partial_z \A_{j}(z,t) &+ \beta_{1j} \partial_t\A_j(z,t) =  2\rmi\gamma_j \A_j(z,t) \\
&\quad\times\int_{-\infty}^t \ud t' R(t-t')  |A_\p(z,t')|^2, \quad j = \s,\i, \notag
\end{align}
where group-velocity dispersion has been neglected. Since we are in the pump-reference frame, $\beta_1$ is the group-slowness relative to the pump. This equation has the solution
\begin{subequations}
\begin{align}
\A_j(z,t) &= \A_j(0,t-\beta_{1j}z) \e^{\rmi\theta_j(z,t)},\quad j = \s,\i \label{field_evolution}\\
 \theta_j(z,t) &= \frac{2\gamma_j}{\beta_{1j}} \int_{t-\beta_{1j} z}^t \ud t' \int_{-\infty}^{t'} \ud t'' \, R(t'-t'')|A_\p(0,t'')|^2,\label{field_phase}
\end{align}
\end{subequations}
which is easily verified by insertion into Eq. \eqref{full_field_eq}. The first integral is due to the walk-off between the pump and the quantum field~\cite{Christensen2016, Bell2015}, while the second term describes the effect of the delayed nonlinear interaction.

\subsection{Joint amplitude for photon pairs}

A convenient way to analyze the state of the generated photon pairs is through consideration of the joint temporal amplitude (JTA) $\JT(t_\s,t_\i)$, defined such that the two-photon part of the state may be written as~\cite{Bell2015}
\begin{equation}
\ket{\psi} =  \int \ud t_\s \int \ud t_\i \JT(t_\s,t_\i) \Ad_\s(t_\s)\Ad_\i(t_\i) \ket{\vac}.
\end{equation} 
The JTA thus describes the joint distribution of temporal states contained in the two-photon state and it holds information about the temporal correlations of the generated photon pairs. The probability of generating a photon pair in a single pump pulse is
\begin{equation}
\label{prob_JTA}
R_\mathrm{pair}  = \int \ud t_\s \int \ud t_\i |\JT(t_\s,t_\i)|^2.
\end{equation}
The joint spectral amplitude (JSA) is often considered instead of the JTA and is the 2D Fourier transform of the JTA
\begin{equation}
\label{2DFourier}
\JS(\omega_\s,\omega_i) = \int \ud t_\s \int \ud t_\i \JT(t_\s,t_\i) \e^{\rmi (\omega_\s t_\s +\omega_\i t_\i)}.
\end{equation}
The unfiltered purity of the heralded photon is determined by the factorability of the JTA or JSA. This is determined by a Schmidt decomposition of the JTA of the form
\begin{equation}
\JT(t_\s,t_\i) = \sum_{n} \lambda_n f_n(t_\s) g_n(t_i),
\end{equation}
where $f_n$ and $g_n$ are Schmidt modes, with the purity being given by~\cite{URen2006}
\begin{equation}
P = \sum_n |\lambda_n|^4 \Big /\left (\sum_n |\lambda_n|^2\right )^2.
\end{equation}
Note that this is only a measure of the degree of the spectral and temporal entanglement of generated photon pairs and not the reduction in state purity resulting from Raman contamination. This is more conveniently expressed through other figures of merit, such as the coincidence-to-accidental ratio. Proposed schemes for linear optical quantum computation relies on interference between photons from different pairs~\cite{Hong1987}, which is limited by state purity~\cite{Mosley2008}, making purity an important property for single-photon sources.

\subsection{Joint amplitude with a non-instantaneous material response}
For the problem under consideration here, we may calculate the JTA from the evolution operator Eq. \eqref{evolution_operator} as
\begin{equation}
\label{JTA_from_U}
\JT(t_\s,t_\i) = \mean{\A_\s(L,t_\s)\A_\i(L,t_\i)\hat{U}(0,L)}.
\end{equation}
Only the three terms in $\hat{U}(0,L)$ with exactly one signal and one idler creation operator contribute to the JTA. Perfoming the calculation (see Appendix \ref{joint_amp_deriv}), we obtain the JTA
\begin{align}
\label{eq:JTA_degenerate}
\JT(t_\s,t_\i) &= \rmi\sqrt{\gamma_\s \gamma_\i}\int_0^L \ud z \W(\tau_\s - \tau_\i) A_\p(0,\tau_\s)A_\p(0,\tau_\i) \notag\\
&\quad\times\exp[\rmi\Phi(z,t_\s,t_\i)],
\end{align}
where we have introduced the function $\W(t)$, which has the spectral form
\begin{align}
\label{W_func}
\W(\omega) &= 1 - f_\R + f_\R \chi_\R'(\Omega - \omega) \notag\\
&\quad+ \rmi f_\R \left [2n_\mathrm{th}(|\Omega - \omega|) + 1\right ] \chi_\R''(|\Omega - \omega|).
\end{align}
Here, the Raman susceptibility $\chi_\R(\omega)$ is the Fourier transform of the Raman response function $h_\R(t)$, with $\chi_\R'(\omega)$ and $\chi_\R''(\omega)$ being its real and imaginary part, respectively.  The function $\W(t)$ determines the temporal separation of the position-dependent creation times, defined by
\begin{align}
\label{collision_coordinates_ph}
\tau_j(z) &= t_j - \beta_{1j}(L - z), \quad j = \s,\i.
\end{align}
These may be interpreted as the creation time of the signal and idler photons, respectively, created at the position $z$ and then detected later at times $t_\s$ and $t_\i$, depending on their propagation speed. 
The phase $\Phi(z,t_\s,t_\i)$ has the form
\begin{align}
\Phi(z,t_\s,t_\i)&= \theta_\p(z,\tau_\s) + \theta_\p(z,\tau_\i)
+\theta_\s(L,t_\s) - \theta_\s(z,\tau_\s)\notag \\
&\quad+ \theta_\i(L,t_\i) - \theta_\i(z,\tau_\i),
\label{JTA_phase}
\end{align}
which has a contribution from accumulated SPM of the pump up till the two creation times $\tau_\s$ and $\tau_\i$ as well as XPM of the produced pair, from the pump, since the time of creation. The phases are given in Eqs. \eqref{pump_phase} and \eqref{field_phase}. \newline
The expression Eq. \eqref{eq:JTA_degenerate} is a very general expression for the JTA in the degenerately pumped scheme, but the integral is fairly intractable. Simpler expressions can be obtained in the limit where the pump pulse is temporally much longer than the time-scale of the function $\W(t)$, defined in Eq. \eqref{W_func}. In this case, we may retain only the DC component and approximate $\W(t) = \W(\omega = 0)\delta(t)$, in which case the creation times coincide: $\tau_\s = \tau_\i = t_\c$ and the JTA takes the simpler form
\begin{align}
\JT_\mathrm{long}(t_\s,t_\i) &=  \frac{\rmi\sqrt{\gamma_\s \gamma_\i}}{|\beta_{1\s}-\beta_{1\i}|}A_\p^2(0,t_\c)\big \{1 - f_\R + f_\R \chi_\R'(\Omega) \notag\\
&\quad +  \rmi f_\R\chi_\R''(\Omega)[2n_{\mathrm{th}}(\Omega)+1]\big \}\notag \\
&\quad\times \exp(\rmi\Phi(t_\s,t_\i))\Theta(z_\c)\Theta(L-z_\c), \label{JTA_long}
\end{align}
where $\Theta$ is the Heaviside step function and the collision coordinates are defined to be
\begin{align}
\label{collision_coordinates}
z_\c &= L - \frac{t_\s - t_\i}{\beta_{1\s} - \beta_{1\i}},    &  t_\c &= \frac{\beta_{1\s} t_\i - \beta_{1\i}t_\s}{\beta_{1\s} - \beta_{1\i}}.
\end{align}
These coordinates may be interpreted as the time of creation $t_\c$ of the photon pair, which in this limit is produced simultaneously, at the waveguide position $z_\c$. The phase reduces to
\begin{align}
\Phi(t_\s,t_\i) &= 2\gamma_\p z_\c  |A_\p(0,t_\c)|^2 + \frac{2\gamma_\s}{\beta_{1\s}} \int_{t_\c}^{t_\s} \ud t'|A_\p(t')|^2 \notag\\
&\quad+\frac{2\gamma_\i}{\beta_{1\i}} \int_{t_\c}^{t_\i} \ud t'  |A_\p(t')|^2.
\end{align}
In the case $f_\R = 0$, with no phononic contribution to the nonlinearity, the JTA Eq. \eqref{JTA_long} reduces to a previously known result~\cite{Bell2015}. The main feature of this expression compared to previous work is the dependence of the overall amplitude on the frequency separation $\Omega$ and the waveguide temperature $T$, through its impact on the phonon population.

\subsection{Photon statistics in the long-pulse limit}

\begin{figure}[ht]
\centering
\includegraphics[scale = 1]{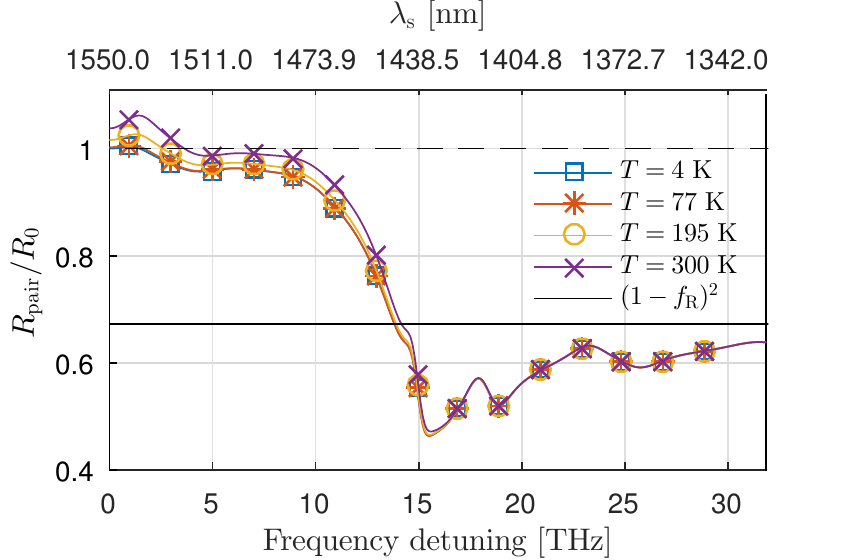}
\caption{The generation probability of photon pairs, normalized to the case of $f_\R = 0$, as a function of linear frequency detuning. The wavelength axis is shown for a $1550$ nm pump. The dashed line marks $r = 1$.}
\label{Raman_ratio}
\end{figure}

The pre-filtering probability $R_\mathrm{pair}$ of generating a pair, is found by integration, which is most easily performed by a change of variables $(t_\s, t_\i) \to (t_\c,z_\c)$, resulting in a Jacobian of $|\beta_{1\s}-\beta_{1\i}|$. Performing the integration yields
\begin{align}
R_\mathrm{pair} &=  \frac{\gamma_\s \gamma_\i L \int \ud t |A_\p(0,t)|^4}{|\beta_{1\s} - \beta_{1\i}|} \big \{[1 - f_\R + f_\R \chi_\R'(\Omega)]^2 \notag \\
&\quad +  f_\R^2\chi_\R''(\Omega)^2[2n_{\mathrm{th}}(\Omega)+1]^2\big \}  .
\end{align}
This result is clearly unphysical in the limit $\beta_{1\s} = \beta_{1\i}$. This is due to the fact that, when higher-order dispersion is neglected, phase-matching is achieved at all frequencies in this limit. Without filtering, this results in an unbounded generation rate (limited only by pump depletion which is also neglected here). For a silica fiber, the Raman response is well-represented by a superposition of 13 simple oscillators~\cite{Hollenbeck2002}. This model is used exclusively in this work. We plot $R_\mathrm{pair}/R_0$ where $R_0$ is the number of produced pairs with a purely electronic response, $f_\R = 0$. This plot is shown in Fig. \ref{Raman_ratio} for four different waveguide temperatures (liquid helium, liquid nitrogen, dry ice and room temperature).
The figure shows that, while detuning the signal/idler beyond the Raman spectrum significantly reduces Raman noise, it can also decrease pair-production efficiency to less than half of its maximum value, and with a far detuning limit of $(1-f_\R)^2 \approx 0.67$.  The temperature dependence of the pair production is seen to be fairly small and cooling the fiber with dry ice (195 K), liquid nitrogen (77 K) or liquid helium (4 K), in order to eliminate noise from single Raman photons, does not significantly decrease the pair-production probability. This is because the FWM-efficiency is dominated by the behavior of the real part of the Raman susceptibility. In other waveguides constructed from different materials, the phononic contribution may be larger than in silica and this effect could be more significant and relevant even for longer pulses. \par
\begin{figure}[ht]
\centering
\includegraphics[scale=1]{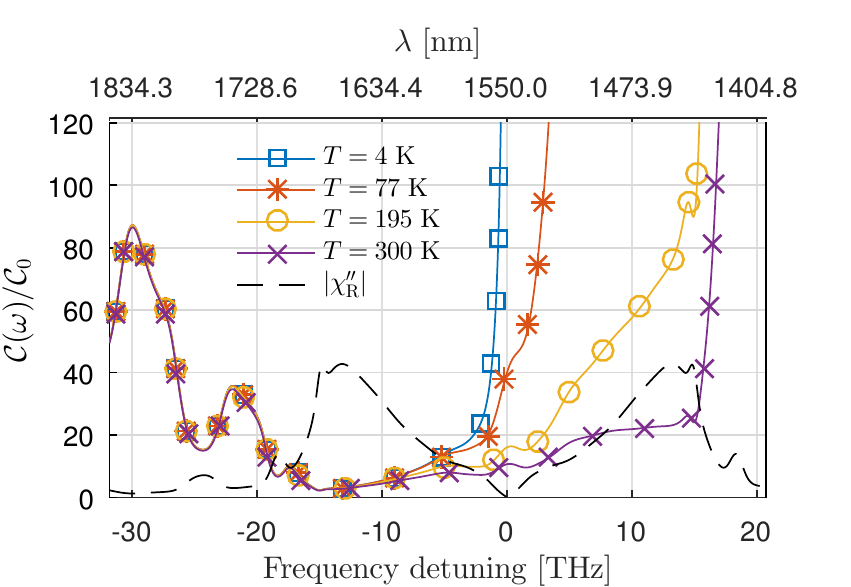}
\caption{The value of $C(\omega,T)$ as a function of linear frequency detuning for three different temperatures with a pump positioned at $\omega = 0$ or $\lambda = 1550$ nm. The dashed line represents the Raman susceptibility.}
\label{fig:C_both}
\end{figure}

The ratio of produced photon pairs to the number of produced single SpRS photons within $\Delta \omega$ of the signal wavelength is given by $\mathcal{C}(\omega) = R_\mathrm{pair}(\omega)/R_\mathrm{R}(\omega)$. The generation probability $R_\R(\omega)$ of single Raman photons is given by Eq. \eqref{eq:prob_SpRS} in Appendix \ref{app:Raman}:
\begin{equation}
R_\mathrm{R}(\omega) = \frac{1}{\pi} \gamma(\omega) f_\R E_\p \Delta\omega \chi_\R''(|\omega|)[n_\mathrm{th}(|\omega|) + \Theta(-\omega)]L,
\end{equation}
where $E_\p$ is the pump pulse energy. This ratio is shown in Fig. \ref{fig:C_both} in units of
\begin{equation}
\mathcal{C}_0 = \frac{\pi\gamma}{\Delta\omega E_\p|\beta_{1\s}-\beta_{1\i}|} \int \ud t |A_\p(0,t)|^4.
\end{equation}
On the anti-Stokes side of the pump (negative frequencies) a strong temperature dependence is present, while on the Stokes side (positive frequencies) a temperature-dependence is only seen for low pump-idler frequency separations. This temperature-dependence is the reason waveguide cooling is often used to suppress SpRS-noise in degenerate copolarized SpFWM when SpFWM occurs within tens of nanometers of the pump~\cite{Takesue2005}. As expected, the Raman resonance peak should be avoided to maximize the ratio of pairs to single photons. Because the FWM efficiency drops for large detunings, since the phononic contribution vanishes, a larger detuning is necessary to achieve a better value of $C(\omega,T)$ than expected by simply looking at the Raman gain spectrum. \par
\begin{figure}[ht]
\centering
\includegraphics[scale=1]{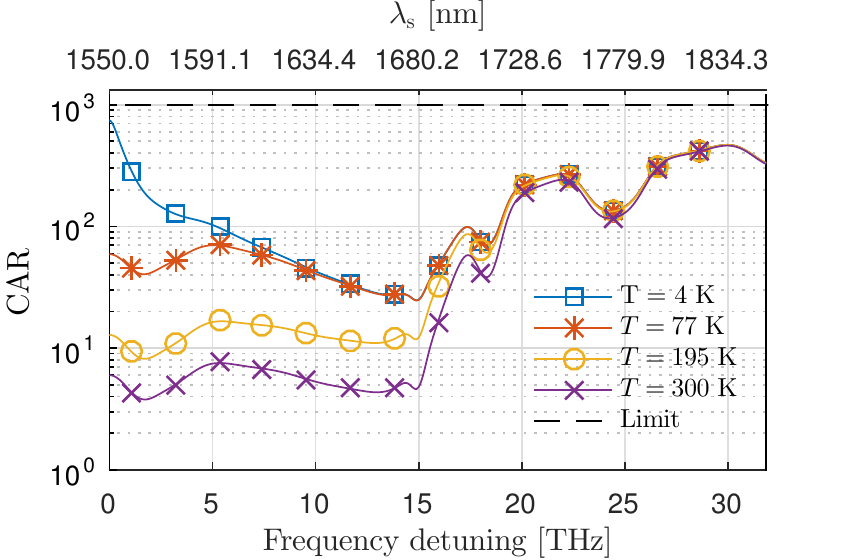}
\caption{The coincidence-to-accidental ratio of a pair-source based on degenerate SpFWM as a function of linear frequency detuning. The wavelength axis assumes a $1550 \, \mathrm{nm}$ pump wavelength.}
\label{fig:CAR}
\end{figure}

Lastly, Fig. \ref{fig:CAR} shows the coincidence-to-accidental ratio of photon pairs, given by
\begin{equation}
\mathrm{CAR} = \frac{R_\mathrm{pair}}{[R_\mathrm{pair} + R_{\R}(\Omega)][R_\mathrm{pair} + R_{\R}(-\Omega)]},
\end{equation}
where the detector dark-count accidents have been neglected. The CAR is shown for a Gaussian pump-input $A_\p(0,t) = \sqrt{P_\p}\exp(-t^2/2T_\p^2)$ with a pulse duration of $T_\p = 1$ ps and a peak power such that the generation probability is fixed at $R_\mathrm{pair} = 0.001$ and a $\Delta\omega$ corresponding to 1 nm at the pump wavelength. For the nonlinearities we use a silica fiber example with $\gamma = 2.0 \, \mathrm{W^{-1} \, km^{-1}}$ and the Raman response for silica.
We note that, as is a well-known experimental fact~\cite{Dyer2009}, the CAR can be significantly increased close to the pump by cooling a silica waveguide, with cooling to $T = 4\, \mathrm{K}$ getting very close to the multi-pair limit $\mathrm{CAR} = 1/R_\mathrm{pair}$ (neglecting dark counts). At larger frequency separations the impact on the CAR is much less significant due to the temperature-independent component of the Stokes scattering. Lastly, we note that this model can, in a straight-forward way, be extended to dual pump configurations, by selecting the appropriate response functions for the various interactions in the evolution operator.

\section{Numerical model} \label{sec:numeric}
\subsection{Split-step scheme for the joint amplitude}

\begin{figure}[ht]
\centering
\def\svgwidth{1\linewidth}
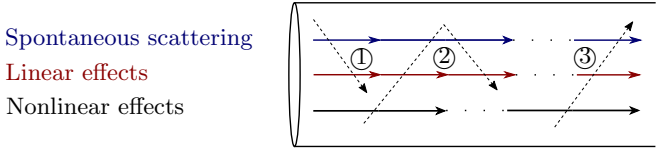
\caption{Application of steps in the developed $\mathcal{O}(h^3)$ split-step scheme with color indicating the type of effect. The dashed arrows indicate the order in which the steps should be applied. The numbers indicate \textcircled{1} the  initialization part, \textcircled{2} the repeating part and \textcircled{3} the final part.}
\label{fig:split_step}
\end{figure}

In this section we describe a numerical scheme for propagating the joint state of a photon pair through a waveguide. It is based on the same idea as the well-known split-step scheme~\cite{Agrawal2013}, which has previously been used to propagate joint states~\cite{Bell2015}. We here present a very generally applicable scheme which can be used for modelling a wide variety of systems. The spatio-temporal evolution of the JTA satisfies an operator differential equation of the form
\begin{subequations}
\begin{align}
\label{formal_equation}
\frac{\ud \JT (z)}{\ud z} &= \hat{f}(z)\JT (z) + \hat{g}(z), \\
\JT(0) &= 0,
\end{align}
\end{subequations}
where $\hat{f}$ represents phase-modulating and dispersive effects (linear and nonlinear) and $\hat{g}$ represents spontaneous scattering effects. The nonlinear (phase-modulating) effects have an exact solution in the time domain, while the linear (dispersive) effects have an exact solution in the frequency-domain. The spontaneous step can be solved exactly in both domains. To develop the split-step scheme, we note that Eq. \eqref{formal_equation} has the formal solution
\begin{align}
\JT(z + h) &= \exp\left (\int_z^{z+h} \ud z' \hat{f}(z')\right ) \Bigg [
\JT(z)  \\
&\quad+\int_z^{z+h} \ud z' \exp\left (-\int_{z}^{z'} \ud z'' \hat{f}(z'')\right ) \hat{g}(z')\Bigg ]\notag .
\end{align}
Invoking the trapezoidal rule to approximate the second integral gives
\begin{align}
\JT(z + h) &=
 \frac{h}{2}\hat{g}(z+h) +\exp\left (\int_z^{z+h} \ud z' \hat{f}(z')\right ) \notag \\
 &\quad\times \left [ \frac{h}{2}\hat{g}(z)+ \JT(z)\right ]+\mathcal{O}(h^3).
\end{align}
This shows that a split-step scheme with a local step-error of order $h^3$ can be constructed by applying half a step of the spontaneous emission effects at position $z$, then apply the usual split-step operations of phase-modulation and dispersion and then the second half of the spontaneous emission at position $z+h$. The linear and nonlinear operation should be applied symmetrically, i.e. half a linear step, followed by a full nonlinear step and ended with another linear half-step. Thus, to have a $\mathcal{O}(h^3)$ split-step scheme, the application of steps should follow the structure illustrated in Fig. \ref{fig:split_step}.

Compared to a naive application of sequentially full steps, which gives an error of $\mathcal{O}(h^2)$, only the linear step, which is the least computationally expensive, needs to be applied twice as many times. Thus, using the order of applications of steps as depicted in Fig. \ref{fig:split_step}, gives an improvement in computation speed for a given required accuracy.\par
By using Eq. \eqref{JTA_from_U}, the spatial evolution of the JTA is given by
\begin{align}
\label{diff_A}
\frac{\partial \JT(z,t_\s,t_\i)}{\partial z} &= \mean{\A_\s(z,t_\s)\A_\i(z,t_\i)\frac{\ud \hat{U}(0,z)}{\ud z}}  \notag\\
&\quad+ \mean{\frac{\partial[\A_\s(z,t_\s)\A_\i(z,t_\i)]}{\partial z}\hat{U}}.
\end{align}
The first term in Eq. \eqref{diff_A} describes spontaneous scattering and is independent of $\JT$. It is evaluated similarly to the calculation in Appendix \ref{joint_amp_deriv}. The second term describes dispersive and phase-modulating effects and is proportional to $\JT$. It is evaluated using the evolution equation of the fields under both group-velocity dispersion and phase-modulation. This results in the following evolution equations involving both the JTA and the JSA:
\begin{subequations}
\begin{align}
\left (\frac{\ud \JS}{\ud z}\right )_{\mathrm{Sp}}
&= \rmi\sqrt{\gamma_\s \gamma_\i} \int_{-\infty}^{\infty} \ud\omega \W(\omega)\notag \\
&\quad\times A_\p(z,\omega_\s + \omega) A_\p(z,\omega_\i - \omega)  \\
\left (\frac{\ud \JS}{\ud z}\right )_{\mathrm{L}} &= \rmi \left(\sum_{n=1}^\infty\beta_{n\s} \frac{\omega_\s^n}{n!} + \sum_{n=1}^\infty \beta_{n\i}\frac{\omega_\i^n}{n!}\right ) \JS(z,\omega_\s,\omega_\i), \\
\left (\frac{\ud \JT}{\ud z}\right )_{\mathrm{NL}} &=
2\rmi \JT(z,t_\s,t_\i) \int_{0}^\infty \ud t' R(t') \\
&\quad\times(\gamma_\s |A_\p(z,t_\s-t')|^2 + \gamma_\i |A_\p(z,t_\i-t')|^2),\notag
\end{align}
\end{subequations}
where the 2D Fourier transform used to go from a temporal to a spectral description is given by Eq. \eqref{2DFourier} and we have suppressed the temporal/spectral dependence of the JTA/JSA. Solving these equations and invoking the trapezoidal rule, consistent with our $\mathcal{O}(h^3)$-scheme, yields the following steps:
\begin{widetext}
\begin{subequations}
\begin{align}
\text{(SpS)} \quad \JS(z+h,\omega_\s,\omega_\i) &= \JS(z,\omega_\s,\omega_\i) + \rmi\frac{h}{2}\sqrt{\gamma_\s \gamma_\i} \int_{-\infty}^\infty \ud \omega \W(\omega) \Big [A_\p(z,\omega_\s + \omega)A_\p(z,\omega_\i - \omega) \\
&\qquad+ A_\p(z+h,\omega_\s + \omega) A_\p(z+h,\omega_\i - \omega)\Big ] + \mathcal{O}(h^3) \notag\\
\text{(linear)} \quad \JS(z+h,\omega_\s,\omega_\i) &= \JS(z,\omega_\s,\omega_\i) \exp\left [ \rmi \left(\sum_{n=1}^\infty\beta_{n\s} \frac{\omega_\s^n}{n!} + \sum_{n=1}^\infty \beta_{n\i}\frac{\omega_\i^n}{n!}\right )h \right ],\\
\text{(nonlinear)} \quad \JT(z+h,t_\s,t_\i) &=  \JT(z,t_\s,t_\i) \exp \Bigg[\rmi h \int_{0}^\infty \ud t' R(t') \Big(\gamma_\s |A_\p(z,t_\s-t')|^2 \\
&\quad +\gamma_\i |A_\p(z,t_\i-t')|^2 +\gamma_\s |A_\p(z+h,t_\s-t')|^2 +\gamma_\i |A_\p(z+h,t_\i-t')|^2\Big)\Bigg ]+ \mathcal{O}(h^3).\notag
\end{align}
\end{subequations}
\end{widetext}
Since knowledge of the pump field at $z + h$ is required for two of the steps, the pump field must be kept one step ahead of the JTA by a regular split-step method, based on the pump-evolution equation. The time integral in the nonlinear step can be carried out as a convolution between the pump envelope and the time-response $R(t)$ while the frequency-integral in the spontaneous scattering-step can be calculated as a 2D convolution between $f(\omega_1,\omega_2) = A_\p(z,\omega_1)A_\p(z,\omega_2) + A_\p(z+h,\omega_1)A_\p(z+h,\omega_2)$ and $g(\omega_1,\omega_2) = \delta(\omega_1 + \omega_2)\W(\omega_1)$. If it is known that the Raman response function can be neglected, one can simply set $\W(\omega) = 1$ (before the Raman peak) or $\W(\omega) = 1-f_\R$ (after the Raman peak) and $R(t) = \delta(t)$. \newline
Extending these equations to the non-degenerate case with two pumps labelled by $\p$ and $\q$ is straight-forward: The spontaneous step gets two contributions of the type $A_\p A_\q$ with their arguments switched and weighted by appropriate response functions, depending on the polarization of pumps and generated fields. The linear step is unchanged and the nonlinear step should include the relevant phase-modulating effects, again depending on relative polarizations.

\subsection{Numerical results on the impact of Raman scattering on photon purity}

The generality of the numerical model allows for analysis of the joint state of photon pairs generated in any SpFWM process with knowledge of the nonlinear response of the material. One feature which has not previously been considered is the impact of the slower phononic response compared to the nearly instantaneous electronic response in silica fibers. For short pump pulses we would expect this to have an impact on the state of generated pairs. We consider a Gaussian pump input of the form
\begin{equation}
A_\p(0,t) = \sqrt{P_\p} \exp\left (-\frac{t^2}{2T_\p^2}\right ),
\end{equation}
and the symmetric scheme for producing single photons of high pre-filtering purity~\cite{Smith2009}. In this scheme, the waveguide is designed so that $\beta_{1\s} = - \beta_{1\i}$ and the waveguide length $L$ chosen to maximize purity. For simplicity, we again use the parameters for a simple silica fiber. \par
\begin{figure}
\centering
\includegraphics[scale = 1.0]{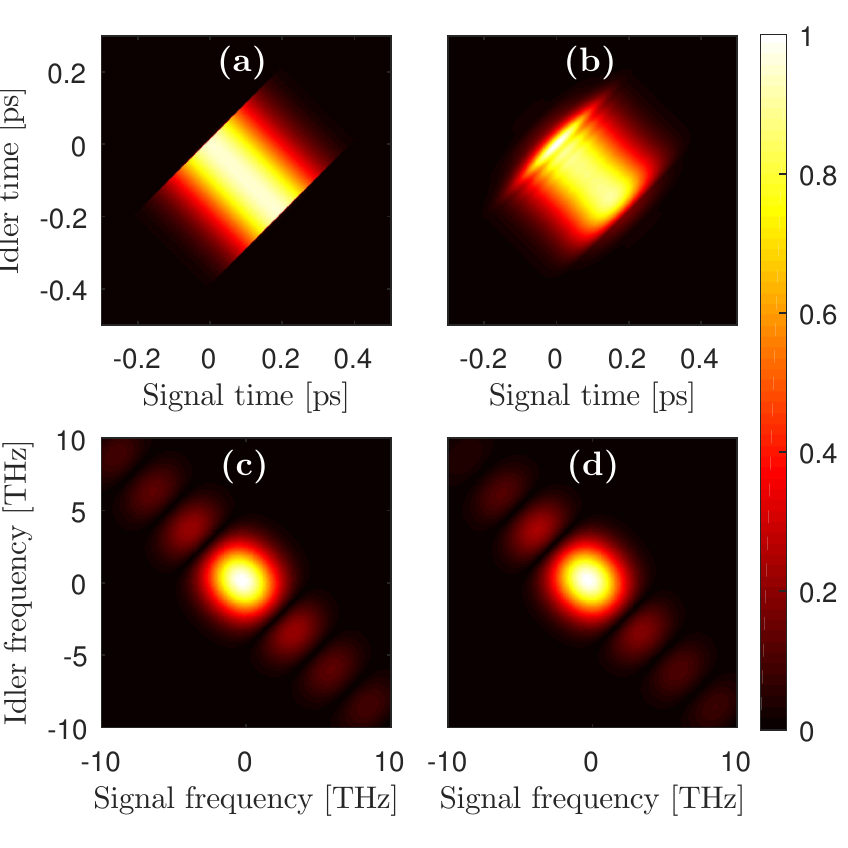}
\caption{The absolute of the joint temporal amplitude with \textbf{(a)} a pure electronic response and \textbf{(b)} a pure phononic response. The absolute value of the joint spectral amplitude with \textbf{(c)} a pure electronic response and \textbf{(d)} a pure phononic response.}
\label{fig:JA_Raman}
\end{figure}

To show the physical impact of the finite response time on the JTA, consider Fig. \ref{fig:JA_Raman}(a)-(b) with respectively $f_\R = 0$ and $f_\R = 1$, for illustrative purposes. Both figures are for a short pump pulse of duration $T_\p = 0.1 \, \mathrm{ps}$ and with $|\beta_{1\s}L/T_\p| = 2$, resulting in nearly maximal purity for this scheme, with an angular frequency detuning of $\Omega = 60 \times 10^{12} \, \mathrm{s}^{-1}$ (corresponding to $9.5 \, \mathrm{THz}$). 
The JTA in the case $f_\R = 0$ has the characteristic hard edges which manifests as the ripples at the sides of the main peak in the JSA. These hard edges correspond to perfect temporal information about the heralded photon after detecting a herald produced at either fiber end. Hence, these edges, or the corresponding ripples in the JSA, are a source of correlation in the photon pair, leading to a purity of $P = 0.81$. The purity can be increased by spectral filtering, but avoiding filtering, especially near the single-photon wavelengths, is desirable to achieve higher brightness and lower error-rates. \newline
The JTA in the case $f_\R = 1$ has much smoother edges, and hence less pronounced ripples in the JSA, compared to the instantaneous response case. When the finite phononic contribution is included, exact temporal information from the fiber ends are no longer obtained by detection of the herald, since the heralded photon could have been produced both temporally earlier or later than the herald. This reduces correlation and leads to a higher pre-filtering purity of $P = 0.85$. The purities are determined with a singular-value decomposition in Matlab. The two edges in the phonon-dominated JTA look different with the one from the beginning of the fiber (lower right edge) varying faster then the one corresponding to the end of the fiber (upper left edge).
This is caused by interference due to the spectral shape of the Raman response: The function $\W(t)$ has a faster varying phase for $t > 0$ than for $t < 0$ yielding more averaging for cases where a low-frequency photon (a signal photon in this case) is produced first. This causes the ripples at the beginning of the fiber (which are caused for signal photons being produced first) to be less pronounced.\par
\begin{figure}[ht]
\centering
\includegraphics[scale=1]{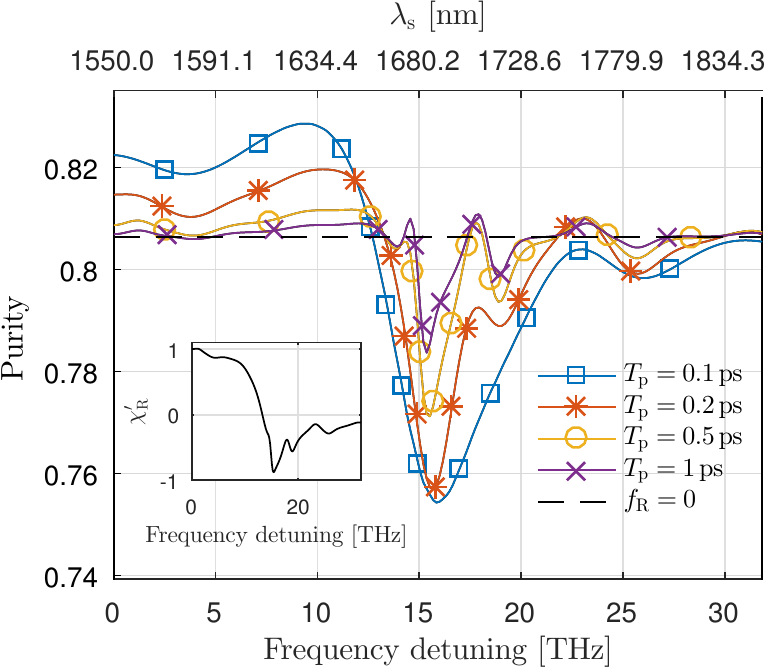}
\caption{The purity of the heralded single-photon as a function of the linear frequency detuning for different pump durations. The wavelength axis assumes a pump wavelength of $1550 \, \mathrm{nm}$. The inset shows the real part of the Raman susceptibility for reference.}
\label{fig:purity_omega}
\end{figure}

Figure \ref{fig:purity_omega} shows the heralded pre-filtering purity as a function of frequency separation for a silica fiber with $f_\R = 0.18$, with four different pump pulse durations in the low pair-production regime.
We see a frequency-dependence of the purity which can be quite significant for short pump pulses for certain frequency separations. The frequency-dependence shows some similarity to the real part of the Raman susceptibility since its magnitude largely determines the relative contribution to pair-generation from phonons. Interestingly, the resulting change in purity can be both positive and negative compared to the case without any phononic contribution, with a very significant decrease in purity just beyond the Raman peak. This is likely caused by the more rapid oscillations of the Raman response introducing correlations and counteracting the beneficial contribution from the smoothening.

\section{Conclusion}

We have developed a mathematical model for the state of photon pairs generated by spontaneous four-wave mixing in the case of a pulsed pump, which includes the full, time-dependent, nonlinear response. Closed form expressions were found in the long pulse limit for the joint amplitude and the generation probability of photon pairs, depending on the frequency separations in the setup and the waveguide temperature. It was found that higher-order Raman scattering events gives the pair-production process a temperature dependence and that, according to the model, for a silica fiber, generation rates depend only weakly on waveguide temperature. The pair-generation probability was however found to depend strongly on frequency separation, with less than half of the expected probability from a purely electronic analysis, for some detunings. \par
In addition, we presented a numerical, symmetrical split-step scheme, to propagate the photon-pair state along the waveguide, where all effects, such as higher-order dispersion, nonlinear phase modulation and a non-instantaneous response can be included at once. This model was used to demonstrate the impact of the finite phononic response time on the purity of heralded single photons produced in a silica fiber, and a significant effect was observed for pulse durations shorter than roughly 1 ps (2.4 ps FWHM) for frequency detunings less than 30 THz. This proves that in addition to noise contamination, Raman scattering can alter the spectral correlations of generated photon pairs. In this work, the numerical model was used to study the impact of a non-instantaneous nonlinear response, but it has general applications in modelling of realistic systems and may be of help in designing such realistic systems, by including all effects of interest and yielding a realistic prediction of produced photon states.

\section*{Acknowledgement}

This work was supported by the Danish Council for Independent Research (DFF) (4184-00433).

\appendix

\section{Hamiltonians \label{app:el_Hamiltonian}}
\subsection{Electronic Hamiltonian}

We take as our starting point the nonlinear interaction energy~\cite{Carter1987}
\begin{equation}
\label{H_initial}
\mathcal{\hat{H}}^{(\e)}_\mathrm{int}(t)= -\frac{\epsilon_0 \chi_\e^{(3)}}{4} \int \ud^3 \vec{r} \normord{|\vec{\hat{E}}|^4},
\end{equation}
where $\chi^{(3)}_\e$ is the part of the relevant component of the nonlinear tensor that stems from electronic interactions, which are assumed to be instantaneous. For simplicity, we derive the Hamiltonians under the assumption that all fields exist in a single spatial mode and that all fields are at approximately the same frequency, such that we take $\omega_{\s 0} \approx \omega_{\p 0} \approx \omega_{\i 0}$, in the sense that $|\omega_{\s 0} - \omega_{\i 0}| \ll \omega_{\p 0}$, in all prefactors. We take $\vec{\hat{E}} = \vec{E}_\p + \vec{\hat{E}}_\s + \vec{\hat{E}}_\i$, with fields given in Eqs. \eqref{pump_field} and \eqref{signal_field}, repeated here for convenience
\begin{align*}
\vec{\hat{E}}_j(z,t) &= \frac{1}{2}\vec{e}F(x,y)\e^{\rmi(\beta_{j 0} - \omega_{j 0}t)} \sqrt{ \frac{2\hbar \omega_{\p 0}}{n_\p \epsilon_0  c }} \A_j(z,t) + \HC, \\
\vec{E}_\p(z,t) &= \frac{1}{2}\vec{e} F(x,y)  \sqrt{\frac{2}{n_\p\epsilon_0 c }}  A_\p(z,t)\e^{\rmi(\beta_{0\p}z - \omega_{0\p}t)} +\cc,
\end{align*}
where $A_\p$ is a classical pump field with units $\sqrt{\mathrm{W}}$ and $\A_j, j = \s,\i$, are quantum fields with units $\mathrm{s}^{-1/2}$. For convenience, we suppress all integral limits in the following, with the understanding that integrals over longitudinal waveguide position are from $0$ to $L$ and time integrals over all time. Inserting the total field into Eq. \eqref{H_initial}, gives
\begin{align}
\label{H_intermed}
\mathcal{\hat{H}}^{(\e)}_\mathrm{int}(t) &= -\frac{3\hbar\omega_{\p 0}\chi_\e^{(3)}}{8 c^2 \epsilon_0 n_\p^2 A_\mathrm{eff}} \int \ud z A_\p^2(z,t) \Ad_{\s,\i}(z,t)^2 \e^{2\rmi \theta_\p}\notag \\
&\quad+ \HC,
\end{align}
where we introduced the total field operator
\begin{equation}
\A_{\s,\i}(z,t) =\A_\s(z,t)\e^{\rmi\theta_\s(z,t)} + \A_\i(z,t)\e^{\rmi \theta_\i(z,t)},
\end{equation}
with $\theta_j(z,t) = \beta_{0j}z - \omega_{0j}t$ and the effective area
\begin{equation}
A_\mathrm{eff} = \left (\iint \ud x \ud y |F(x,y)|^4 \right )^{-1}.
\end{equation}
Defining the electronic nonlinear parameter as~\cite{Agrawal2013}
\begin{equation}
\label{nonlinear_coeff}
\gamma_\e = \frac{3\chi_\e^{(3)}\omega_{\p 0}}{4 \epsilon_0 c^2 n_\p^2 A_\mathrm{eff}},
\end{equation}
allows us to write the Hamiltonian in the simpler form
\begin{equation}
\label{H_intermed2}
\mathcal{\hat{H}}^{(\e)}_\mathrm{int}(t) = -\frac{\hbar\gamma_\e}{2} \int \ud z A_\p^2(z,t) \Ad(z,t)^2 \e^{2\rmi\theta_\p}+ \HC
\end{equation}

\subsection{Generalization to non-instantaneous response}

We seek a Hamiltonian that generates the following nonlinear Heisenberg equation for the slowly-varying field envelope $\a(z,t)$~\cite{Lin2007}
\begin{align}
\label{Heisenberg}
\partial_z \a(z,t) &= \rmi \gamma A_\p(z,t)\int \ud t' R(t-t') A_\p(z,t') \ad(z,t')\notag\\
&\quad+ \rmi \M(z,t)A_\p(z,t),
\end{align}
where $R(t) = (1-f_\R) \delta(t) + f_\R h_\R(t)$ is the nonlinear response function with $f_\R$ being the fraction of the nonlinear response from the phonons (the remaining coming from the electrons) and $h_\R(t)$ the normalized Raman response. Note that the electronic nonlinear parameter is thus $\gamma_\e = (1-f_\R)\gamma$, where $\gamma$ is the total nonlinear parameter. The noise operator $\M$ is introduced in Sec. \ref{H_phon}. In the frame moving along the pump at its group velocity we may change from a Hamiltonian governing time evolution to a Hamiltonian governing spatial evolution, such that the evolution operator is unchanged:
\begin{equation}
\exp\left (-\frac{\rmi}{\hbar} \int_{-\infty}^\infty \ud t \mathcal{\hat{H}}_\mathrm{int}(t) \right ) = \exp\left (\rmi \int_0^L \ud z \H_\mathrm{int}(z)\right ),
\end{equation}
where temporal evolution from a very early time, before the pulse enters the waveguide, to a very late time, long after the pulse has left the waveguide, is replaced by evolution of the pulse through the waveguide. Doing this, we assume that time-ordering corrections can be neglected~\cite{Quesada2014}, which is only valid in the low-gain regime~\cite{Christ2013a}. Thus, Eq. \eqref{H_intermed2} corresponds to
\begin{align}
\H^{(\e)}_\mathrm{int}(z) &= -\frac{\gamma_\e}{2} \int \ud t A_\p^2(z,t) \left [\Ad(z,t) + \A(z,t)\right ] \notag \\
&\quad \times\Ad(z,t) + \HC,
\end{align}
Note that this is strictly speaking a momentum operator since it is a generator of spatial translations. We now generalize this in a straight-forward way to include a non-instantaneous response
\begin{align}
\label{H_nonins}
\H_\mathrm{int}^{(1)}(z)& = \frac{\gamma}{2} \iint \ud t_1 \ud t_2 R(t_2 - t_1) A_\p(z,t_1) A_\p(z,t_2) \notag \\
&\qquad \times(\A + \Ad) \Ad(z,t) + \HC,
\end{align}
and show that this Hamiltonian generates the Heisenberg equations for the fields. To do this, introduce $\hat{f} = \Ad + \A$ and note that $[\hat{f}(z,t), \hat{f}(z,t')] = 0$ and $[\hat{f}(z,t), \Ad(z,t')] = 2\delta(t-t')$. Using this, we find the Heisenberg equation of motion for the field:
\begin{align}
\partial_z \hat{f} &=  \rmi \left [\hat{f}(z,t), \H_\mathrm{int}^{(1)}(z)\right ] \\
&= \rmi\gamma\int \ud t_1 R(t - t_1) A_\p(z,t_1) A_\p(z,t) \notag \\
&\qquad \times\Big[\A_\s(z,t_1)\e^{\rmi (2\theta_\p -\theta_\s)} + \Ad_\s(z,t_1)\e^{-\rmi (2\theta_\p -\theta_\s)}  \notag\\
&\qqquad+ \A_\i(z,t_1)\e^{\rmi(2\theta_\p -\theta_\i)}+ \Ad_\i(z,t_1)\e^{-\rmi (2\theta_\p -\theta_\i)}\Big]\notag \\
&\quad+ \HC
\end{align}
As usual in this kind of equation we compare the phase-matched terms to find the evolution of e.g. the signal field:
\begin{align}
\partial_z \A_\s(z,t) &= \rmi\gamma A_\p(z,t)\int \ud t' R(t - t') A_\p(z,t')\notag \\
&\quad\times\Ad_\i(z,t') \e^{-i\Omega (t-t')},
\end{align}
which indeed agrees with the Heisenberg equation Eq. \eqref{Heisenberg}. Within the slowly-varying envelope approximation, the Hamiltonian Eq. \eqref{H_nonins} can, by expansion of the field $\hat{f}$ and exchange of $t_1$ and $t_2$ in one of the terms, be written in the simpler form
\begin{align}
\label{H_nonins2}
\H_\mathrm{int}^{(1)}(z)& = \gamma\iint \ud t_1 \ud t_2 \RR(t_2 - t_1) A_\p(z,t_1) A_\p(z,t_2) \notag \\
&\qquad \times \Ad_\s(z,t_1) \Ad_\i(z,t_2) \e^{-i\Omega (t_1-t_2)}+ \HC,
\end{align}
where we introduced the function
\begin{equation}
\RR(t) = \frac{1}{2}\left [R(t) + R(-t)\right ],
\end{equation}
that governs the temporal separation of pair-creation events. This Hamiltonian has the desired properties of being Hermitian, being symmetric in the signal and idler fields and reducing to the purely electronic case in the limit $h_\R(t) \to \delta(t)$, as it should.

\subsection{Spontaneous Raman scattering \label{H_phon}}

We use a standard model for the phonon interaction and model the noise background as a continuum of independent and localized harmonic oscillators, with weight $W(\omega)$~\cite{Boivin1994}:
\begin{equation}
\M(z,t) = \int_0^\infty \ud \omega \frac{\sqrt{W(\omega)}}{2\pi} \left [\hat{d}^\dagger_\omega(z) \e^{\rmi\omega t} + \hat{d}_{\omega}(z) \e^{-\rmi\omega t}  \right ],
 \end{equation}
where $\hat{d}^\dagger_\omega(z)$ is the phonon creation operator at waveguide position $z$ and angular frequency $\omega$. These are uncorrelated at different position, to represent local oscillatory excitations, and normalized such that $[\hat{d}_\omega(z), \hat{d}^\dagger_{\omega'}(z')] = \delta(\omega-\omega')\delta(z-z')$. We only consider a material at thermal equilibrium so that they have correlations
\begin{equation}
\label{phonon_operator}
\mean{\hat{d}^\dagger_{\omega'}(z') \hat{d}_\omega(z)}_\mathrm{th} = n_\mathrm{th}(\omega)\delta(\omega-\omega')\delta(z-z'),
\end{equation}
where $n_\mathrm{th}(\omega) = [\exp(\hbar \omega/k_\mathrm{B}T) -1]^{-1}$ is the expected occupation number of states at frequency $\omega$ and waveguide temperature $T$. The nonlinear response, given in Eq. \eqref{NLresponse}, has the spectral form $R(\omega) = 1-f_\R + f_\R [\chi_\R'(\omega) + \rmi \chi_\R''(\omega)]$. The spectral density of phonon modes $W(\omega)$ can be shown to be related to the imaginary part of the nonlinear response by $W(\omega) = 4\pi \mathrm{Im}\{\gamma R(\omega)\}  = 4\pi \gamma  f_\R \chi_\R''(\omega)$~\cite{Boivin1994}. The Hamiltonian generating the relevant evolution term is easily identified as
\begin{align}
\H_\mathrm{int}^{(2)}(z) &= \int \ud t  A_\p(z,t) \M(z,t) \\
&\quad\times\Big [\Ad_\s(z,t)\e^{-\rmi\Omega t} + \Ad_\i(z,t) \e^{\rmi\Omega t}\Big] + \HC\notag
\end{align}
by checking that e.g. $\partial_z \A_\s = i[\A_\s, \H_\mathrm{int}^{(2)}] = \rmi \M A_\p \e^{-\rmi \Omega t}$ where the oscillating exponential comes from the frequency detuning between the pump and signal field since these fields are spectrally centered around their respective central frequencies. For convenience, we introduce the renormalized noise operator
\begin{equation}
\label{noise_operator}
\m(z,t) =\int_0^\infty \ud \omega \sqrt{\frac{f_\R\chi_\R''(\omega)}{\pi}}  \left [\hat{d}^\dagger_\omega(z) \e^{\rmi\omega t} + \hat{d}_{\omega}(z) \e^{-\rmi\omega t}  \right ].
\end{equation}
Introducing different nonlinear coefficients for the pump, signal and idler fields, such that
\begin{equation}
\label{nonlinear_coeff_2}
\gamma_j = \frac{3\chi^{(3)}\omega_{j 0}}{4 \epsilon_0 c^2 n_\p n_j A_\mathrm{eff}}, \quad j =\p, \s, \i,
\end{equation}
the full interaction Hamiltonian is straight-forwardly generalized to
\begin{align}
\H_{\mathrm{int}}(z)  &=\sqrt{\gamma_\s \gamma_\i} \int_{-\infty}^{\infty}\ud t_1\int_{-\infty}^{\infty}\ud t_2  \RR(t_1 - t_2)\notag \\
&\quad\times A_\p(z,t_1)A_\p(z,t_2)  \Ad_\s(z,t_1) \Ad_\i(z,t_2) \e^{-\rmi\Omega (t_1 - t_2)} \notag\\
&+ \int_{-\infty}^{\infty} \ud t  A_\p(z,t) \m(z,t) \\
&\quad\times\Big [\sqrt{\gamma_\s}\Ad_\s(z,t)\e^{-\rmi\Omega t}+\sqrt{\gamma_\i}\Ad_\i(z,t) \e^{\rmi\Omega t}\Big] +\HC \notag,
\end{align}
where a single spatial mode is still assumed. This Hamiltonian describes the spatial photon-state evolution in the interaction picture where the photon operators evolve under the free part of the Hamiltonian governing dispersive and phase-modulating effects, formulated in this work through Heisenberg evolution equations.

\section{Noise correlations\label{app:noise}}

Using the definition Eq. \eqref{noise_operator} of the noise operator and the thermal correlation Eq. \eqref{phonon_operator} of the phonon operators, the noise correlation functions are straight-forward to calculate:
\begin{align}
\langle \m(z_1,t_1) &\m(z_2,t_2)\rangle \notag \\
 &= \frac{f_\R \gamma }{\pi}\int_0^{\infty}\ud \omega\int_0^{\infty}  \ud \omega' \sqrt{\chi_\R''(\omega)\chi_\R''(\omega')} \notag\\
 &\quad\times \left [\mean{\hat{d}^\dagger_{{\omega}}(z_1) \hat{d}_{\omega'}(z_2)}_\mathrm{th}\e^{\rmi (\omega t_1 - \omega' t_2)} \right . \notag\\
 &\qquad + \left . \mean{\hat{d}_{\omega}(z_1) \hat{d}^\dagger_{\omega'}(z_2)}_\mathrm{th}\e^{-\rmi (\omega t_1 - \omega' t_2)} \right ] \\
 &=\frac{f_\R \gamma }{\pi}\delta(z_1 - z_2)\int_0^{\infty} \ud \omega \chi_\R''(\omega)\notag\\
 &\quad\times \left [n_\mathrm{th}(\omega)\e^{\rmi \omega (t_1 - t_2)} \right . \notag\\
 &\qquad + \left . (n_\mathrm{th}(\omega) + 1)\e^{-\rmi \omega( t_1 -  t_2)} \right ]. \notag
\end{align}
We see that this correlation function takes the form
\begin{equation}
\label{noise_correlation_time}
\mean{\m(z_1,t_1) \m(z_2,t_2)} = \delta(z_1 - z_2) \mathcal{F}(t_1 - t_2),
\end{equation}
where the function $\mathcal{F}(t)$ is much more neatly expressed in the frequency domain as
\begin{equation}
\label{F_omega}
\mathcal{F}(\omega) = 2f_\R\chi_\R''(|\omega|) \left [n_\mathrm{th}(|\omega|)+\Theta(\omega)\right ],
\end{equation}
with a Fourier transform as used in Eq. \eqref{2DFourier}. Here, $\Theta$ is the Heaviside step-function.

\section{Derivation of joint amplitude}\label{joint_amp_deriv}

According to Eq. \eqref{JTA_from_U}, the joint temporal amplitude can be calculated from the evolution operator, given in Eq. \eqref{evolution_operator}, as
\begin{equation}
\JT(t_\s,t_\i) = \mean{\A_\s(L,t_\s)\A_\i(L,t_\i) \hat{U}(0,L)},
\end{equation}
and including the three contributing terms in $\hat{U}(0,L)$, gives the expression
%
\begin{align}
\label{JTA_full}
\JT&(t_\s,t_\i) =\rmi \sqrt{\gamma_\s \gamma_\i} \iiint \ud z \ud t_1 \ud t_2 \RR(t_1-t_2)   \notag\\
&\qqquad\quad\times A_\p(z,t_1)A_\p(z,t_2) \e^{-\rmi\Omega (t_1 - t_2)} \notag\\
&\qqquad\quad\times \mean{\A_\s(L,t_\s) \A_\i(L,t_\i)\Ad_\s(z,t_1) \Ad_\i(z,t_2)} \notag \\
&\qqquad- \frac{1}{2} \sqrt{\gamma_\s \gamma_\i}\iiiint \ud z_1 \ud z_2 \ud t_1\ud t_2 \\
&\qqquad \quad \times A_\p(z_1,t_1) A_\p(z_2,t_2) \mean{\m(z_1,t_1) \m(z_2,t_2)} \notag\\
 &\times\Big[\mean{\A_\s(z,t_\s) \A_\i(z,t_\i)\Ad_\s(z_1,t_1)\Ad_\i(z_2,t_2)}\e^{-\rmi\Omega (t_1 - t_2)}\notag \\
 &\qquad+ \mean{\A_\s(z,t_\s) \A_\i(z,t_\i)\Ad_\i(z_1,t_1)\Ad_\s(z_2,t_2)} \e^{\rmi\Omega(t_1 - t_2)}\Big]. \notag
\end{align}
%
Carrying out one of the space-integrals in the second part of Eq. \eqref{JTA_full} by use of the delta-function in Eq. \eqref{noise_correlation_time}, and using the noise correlation Eq. \eqref{noise_correlation_time}, we see that all three terms in this expression are almost identical, and by defining
\begin{equation}
\label{W_t}
\W(t) = \frac{1}{2}\left \{R(t) + R(-t) + \rmi\left[\mathcal{F}(t) + \mathcal{F}(-t) \right]  \right \} \e^{-\rmi \Omega t},
\end{equation}
we may combine them to
\begin{align}
\JT(t_\s,t_\i) &=\rmi \sqrt{\gamma_\s \gamma_\i} \iiint \ud z \ud t_1 \ud t_2 \W(t_1-t_2) \notag \\
&\quad\times \mean{\A_\s(L,t_\s) \A_\i(L,t_\i)\Ad_\s(z,t_1) \Ad_\i(z,t_2)}\notag \\
&\quad\times A_\p(z,t_1) A_\p(z,t_2).
\end{align}
Using Eq. \eqref{field_evolution}, we may evaluate the field correlations as
\begin{align}
\label{field_correlation}
\mean{\A_j(L,t)\Ad_j(z,t')}& = \mean{\A_j(0,t - \beta_{1j} L)\Ad_j(0,t' - \beta_{1j} z)} \notag \\
&\quad\times\e^{\rmi [\theta_j(L,t) - \theta_j(z,t')]}\notag \\
&= \delta(t - t' - \beta_{1j}(L - z))\\
&\quad\times\e^{\rmi [\theta_j(L,t) - \theta_j(z,t')]}, \quad j = \s,\i,\notag
\end{align}
where the field phase is given by Eq. \eqref{field_phase}. We can use the two delta-functions from the field correlations to carry out the two time integrals, with contributions at $t_1 = \tau_\s$ and $t_2 = \tau_\i$, being the solutions to the delta-conditions
\begin{subequations}
\begin{align}
t_\s - t_1 - \beta_{1\s}(L-z) &= 0, \\
t_\i - t_2 - \beta_{1\i}(L-z) &= 0.
\end{align}
\end{subequations}
This gives the final solution
\begin{align}
\JT(t_\s,t_\i) &= \rmi\sqrt{\gamma_\s \gamma_\i}\int_0^L \ud z \W(\tau_\s - \tau_\i) A_\p(0,\tau_\s)A_\p(0,\tau_\i) \notag \\
&\quad\times\exp[\rmi\Phi(z,t_\s,t_\i)],
\end{align}
where the pump phases are included in the overall phase-factor given by
\begin{align}
\Phi(z,t_\s,t_\i)&=  \theta_\p(z,\tau_\i) + \theta_\p(z,\tau_\s)+\theta_\s(L,t_\s)  - \theta_\s(z,\tau_\s)  \notag\\
&\quad + \theta_\i(L,t_\i) - \theta_\i(z,\tau_\i)\notag\\
&= \gamma_\p z \int_0^{\infty} \ud t R(t) |A_\p(0,\tau_\i(z)-t)|^2\notag\\
&\quad+ \gamma_\p z \int_0^{\infty} \ud t R(t) |A_\p(0,\tau_\s(z)-t)|^2\notag \\
&\quad+ \frac{2\gamma_\s}{\beta_{1\s}} \int_{\tau_\s(z)}^{t_\s} \ud t \int_{0}^{\infty} \ud t' \, R(t')|A_\p(t-t')|^2\notag \\
&\quad+\frac{2\gamma_\i}{\beta_{1\i}} \int_{\tau_\i(z)}^{t_\i} \ud t \int_{0}^{\infty} \ud t' \, R(t')|A_\p(t-t')|^2,
\end{align}
which describes pump self-phase accumulated until the time of creation for both signal and idler as well as signal- and idler pump cross-phase accumulated since creation to the end of the waveguide. We note that
\begin{align}
\frac{1}{2}\left [R(\omega) + R(-\omega)\right ] &= \left [1 - f_\R + \frac{f_\R}{2}\{\chi_\R(\omega) +\chi_\R(-\omega)\}\right ]\notag \\
&= 1 - f_\R + f_\R \chi'_\R(\omega),
\end{align}
and similarly for $\mathcal{F}(\omega)$
\begin{align}
\frac{1}{2}\left [\mathcal{F}(\omega) + \mathcal{F}(-\omega)\right ] &= f_\R\chi_\R''(|\omega|) [2n_\mathrm{th}(|\omega|) \\
&\quad+ \Theta(\omega) + \Theta(-\omega) ] \notag\\
&= f_\R\chi_\R''(|\omega|) [2n_\mathrm{th}(|\omega|) + 1],
\end{align}
and conclude that the function $\W(t)$ has the Fourier transform
\begin{align}
\W(\omega) &= 1 - f_\R + f_\R \chi'_\R(\Omega - \omega) \notag\\
&\quad+ \rmi f_\R [2n_\mathrm{th}(|\Omega - \omega|) + 1] \chi_\R''(|\Omega - \omega|).
\end{align}
It is thus not simply $1 - f_\R + f_\R \chi_\R$ as one might naively expect. Since spontaneous Raman scattering is involved through active interaction with the phonon modes, the amplitude becomes temperature dependent since the phonon bath is assumed in thermal equilibrium.

\section{Rate of single Raman photons \label{app:Raman}}

Since Raman scattering events are not coherent, the probability of such events cannot be computed like for the joint amplitude by projecting the state onto a temporal state basis. Instead we must consider the relevant term in $\ip{\psi}{\psi} = \mean{\hat{U}^\dagger \hat{U}}$ with the evolution operator in Eq. \eqref{evolution_operator}. For a single Raman scattering event this is
\begin{align*}
R &= \gamma_\s f_\R \iiiint \ud z_1 \ud z_2\ud t_1  \ud t_2 A_\p^*(z_1,t_1) A_\p(z_2,t_2)  \\
&\quad \times \mean{\m(z_1,t_1)\m(z_2,t_2)}\mean{\A_\s(z_1,t_1)\Ad_\s(z_2,t_2)}.
\end{align*}
Using Eq. \eqref{noise_correlation_time} for the noise correlation and the field correlation Eq. \eqref{field_correlation} gives
\begin{align*}
R &= \gamma_\s f_\R \iiiint \ud z_1 \ud z_2\ud t_1  \ud t_2 A_\p^*(z_1,t_1) A_\p(z_2,t_2)  \\
&\quad \times \delta(z_1 - z_2)\mathcal{F}(t_1 - t_2)\mean{\A_\s(z_1,t_1)\Ad_\s(z_2,t_2)}  \\
&=\gamma_\s f_\R \iiint \ud z_1 \ud t_1  \ud t_2 A_\p^*(z_1,t_1)) A_\p(z_1,t_2) \\
&\quad\times \mathcal{F}(t_1 - t_2)\delta(t_1 - t_2)  \\
&=\gamma_\s f_\R \mathcal{F}(0) \iint \ud z_1  \ud t_1  |A_\p(z_1,t_1)|^2.
\end{align*}
The value of $\mathcal{F}(0)$ is the integral of its Fourier transform in Eq. \eqref{F_omega} and recognizing the last integral as the total energy in a single pulse $E_\p$ (note that $|A_\p(z,t)|$ is independent of $z$), gives
\begin{equation}
\label{prob_single_Raman}
R = \frac{1}{\pi}\gamma_\s f_\R  E_\p L \int \ud \omega \chi_\R''(|\omega|)[n_\mathrm{th}(|\omega| + \Theta(-\omega)].
\end{equation}
For narrow spectral filtering, compared to the Raman spectrum, the probability of generating a Raman photon at frequency $\omega$ is then
\begin{equation}
\label{eq:prob_SpRS}
R_\R(\omega) = \frac{1}{\pi}\gamma_\s f_\R  E_\p \Delta\omega L \chi_\R''(|\omega|)[n_\mathrm{th}(|\omega| + \Theta(-\omega)].
\end{equation}
which is a well-known result~\cite{Brainis2012}. As is always the case for SpRS, the probability of generating a photon on the Stokes side ($\omega < 0$) is larger than the probability of generating an anti-Stokes photon ($\omega > 0$). \newline
For complete consistency we ought to also include second-order spontaneous Raman scattering events in the noise flux as we did when calculating the joint amplitude. There are fundamentally two types of contributions: Two-photon scattering coherent contributions, similar to the one that contributes to SpFWM, where the photons are created at a single waveguide position and events where two Raman photons are independently created at different waveguide positions. Both of these types of terms appear with two photons in either channel or one in each. However, for simplicity we only include the dominant term in this analysis.

\end{document}

%% file: Feynman_diagrams.pdf_tex
\begingroup%
  \makeatletter%
  \providecommand\color[2][]{%
    \errmessage{(Inkscape) Color is used for the text in Inkscape, but the package 'color.sty' is not loaded}%
    \renewcommand\color[2][]{}%
  }%
  \providecommand\transparent[1]{%
    \errmessage{(Inkscape) Transparency is used (non-zero) for the text in Inkscape, but the package 'transparent.sty' is not loaded}%
    \renewcommand\transparent[1]{}%
  }%
  \providecommand\rotatebox[2]{#2}%
  \ifx\svgwidth\undefined%
    \setlength{\unitlength}{527.91573111bp}%
    \ifx\svgscale\undefined%
      \relax%
    \else%
      \setlength{\unitlength}{\unitlength * \real{\svgscale}}%
    \fi%
  \else%
    \setlength{\unitlength}{\svgwidth}%
  \fi%
  \global\let\svgwidth\undefined%
  \global\let\svgscale\undefined%
  \makeatother%
  \begin{picture}(1,0.29838942)%
    \put(0.16870986,0.11612516){\color[rgb]{0,0,0}\makebox(0,0)[lb]{\smash{}}}%
    \put(0,0){\includegraphics[width=\unitlength,page=1]{Feynman_diagrams.pdf}}%
    \put(0.24886594,0.04009148){\color[rgb]{0,0,0}\makebox(0,0)[lb]{\smash{$\omega$}}}%
    \put(0.29268129,0.18331059){\color[rgb]{0,0,0}\rotatebox{31.91023428}{\makebox(0,0)[lb]{\smash{$\Omega$}}}}%
    \put(0.11846979,0.22998483){\color[rgb]{0,0,0}\rotatebox{-32.80724086}{\makebox(0,0)[lb]{\smash{$\omega - \Omega$}}}}%
    \put(0.04523273,0.23016341){\color[rgb]{0,0,0}\makebox(0,0)[lb]{\smash{\textbf{(a)}}}}%
    \put(0,0){\includegraphics[width=\unitlength,page=2]{Feynman_diagrams.pdf}}%
    \put(0.60234419,0.04795644){\color[rgb]{0,0,0}\rotatebox{-32.26476787}{\makebox(0,0)[lb]{\smash{$\omega$}}}}%
    \put(0.43973138,0.03264149){\color[rgb]{0,0,0}\rotatebox{31.56844598}{\makebox(0,0)[lb]{\smash{$\Omega$}}}}%
    \put(0.50645695,0.12083249){\color[rgb]{0,0,0}\rotatebox{90.20608533}{\makebox(0,0)[lb]{\smash{$\omega + \Omega$}}}}%
    \put(0.40063536,0.23016341){\color[rgb]{0,0,0}\makebox(0,0)[lb]{\smash{\textbf{(b)}}}}%
    \put(0.64581264,0.23016341){\color[rgb]{0,0,0}\makebox(0,0)[lb]{\smash{\textbf{(c)}}}}%
    \put(0,0){\includegraphics[width=\unitlength,page=3]{Feynman_diagrams.pdf}}%
    \put(0.82807178,0.05862369){\color[rgb]{0,0,0}\makebox(0,0)[lb]{\smash{$\Omega$}}}%
    \put(0.73447012,0.0287678){\color[rgb]{0,0,0}\rotatebox{51.22805963}{\makebox(0,0)[lb]{\smash{$\omega_1$}}}}%
    \put(0.92531837,0.06260249){\color[rgb]{0,0,0}\rotatebox{-48.94879697}{\makebox(0,0)[lb]{\smash{$\omega_2$}}}}%
    \put(0.69142842,0.1894689){\color[rgb]{0,0,0}\rotatebox{-51.2339422}{\makebox(0,0)[lb]{\smash{$\omega_1-\Omega$}}}}%
    \put(0.91929166,0.10893363){\color[rgb]{0,0,0}\rotatebox{49.87367011}{\makebox(0,0)[lb]{\smash{$\omega_2+\Omega$}}}}%
    \put(0,0){\includegraphics[width=\unitlength,page=4]{Feynman_diagrams.pdf}}%
    \put(0.02014856,0.06999502){\color[rgb]{0,0,0}\rotatebox{90}{\makebox(0,0)[lb]{\smash{Time}}}}%
  \end{picture}%
\endgroup%

%% file: split_step.pdf_tex
\begingroup%
  \makeatletter%
  \providecommand\color[2][]{%
    \errmessage{(Inkscape) Color is used for the text in Inkscape, but the package 'color.sty' is not loaded}%
    \renewcommand\color[2][]{}%
  }%
  \providecommand\transparent[1]{%
    \errmessage{(Inkscape) Transparency is used (non-zero) for the text in Inkscape, but the package 'transparent.sty' is not loaded}%
    \renewcommand\transparent[1]{}%
  }%
  \providecommand\rotatebox[2]{#2}%
  \ifx\svgwidth\undefined%
    \setlength{\unitlength}{361.54952642bp}%
    \ifx\svgscale\undefined%
      \relax%
    \else%
      \setlength{\unitlength}{\unitlength * \real{\svgscale}}%
    \fi%
  \else%
    \setlength{\unitlength}{\svgwidth}%
  \fi%
  \global\let\svgwidth\undefined%
  \global\let\svgscale\undefined%
  \makeatother%
  \begin{picture}(1,0.22329009)%
    \put(0,0){\includegraphics[width=\unitlength,page=1]{split_step.pdf}}%
    \put(-0.00000001,0.1565541){\color[rgb]{0,0,0.45098039}\makebox(0,0)[lb]{\smash{\textcolor{col1}{Spontaneous scattering}}}}%
    \put(0.0005316,0.10385334){\color[rgb]{0.54117647,0,0}\makebox(0,0)[lb]{\smash{\textcolor{col2}{Linear effects}}}}%
    \put(0.0013698,0.05037155){\color[rgb]{0,0,0}\makebox(0,0)[lb]{\smash{\textcolor{col3}{Nonlinear effects}}}}%
    \put(0,0){\includegraphics[width=\unitlength,page=2]{split_step.pdf}}%
    \put(0.52918415,0.12700046){\color[rgb]{0,0,0}\makebox(0,0)[lb]{\smash{\textcircled{1}}}}%
    \put(0.6564978,0.12848212){\color[rgb]{0,0,0}\makebox(0,0)[lb]{\smash{\textcircled{2}}}}%
    \put(0.87282013,0.12771977){\color[rgb]{0,0,0}\makebox(0,0)[lb]{\smash{\textcircled{3}}}}%
    \put(0,0){\includegraphics[width=\unitlength,page=3]{split_step.pdf}}%
  \end{picture}%
\endgroup%